# Reasons underlying certain tendencies in the data on the frequency of codon usage

**Semenov D.A.**

**dasem@mail.ru**

*Abstract. The tendencies described in this work were revealed in the course of examination of adenine and uracil distribution in the mRNA encoding sequence. The study also discusses the usage of codons occupied by the amino acid arginine in the table of the universal genetic code. All of the described tendencies are qualitative, so neither sophisticated methods nor cumbersome calculations are necessary to reveal and interpret them.*

The data used in this study were found in the database (http://www.kazusa.or.jp/codon/). In total, 4500 "species", represented in the database by more than 5000 triplets each, were selected.

*Arginine codons.*

In the table of the universal genetic code, the amino acid arginine is represented by six codons: CGA, CGG, CGU, CGC, AGG, and AGA. Thus, the mutation in the first letter of codons CGA and CGG does not affect the protein primary sequence. This mutation, unpunished by selection, must be much more frequent than the similar mutation in codons CGU and CGC, in which arginine is replaced by serine.

This idea was confirmed by our analysis of the frequency of occurrence of codons. The value calculated from the data on the frequency of occurrence, using the formula (CGA+CGG-CGU-CGC), was found to be less than 0 for 80% of the examined genomes. The other 20% were mainly represented by genomes based on specific dialects of the genetic code, where codons AGG and AGA do not correspond to arginine.

Thus, due to the oxidative guanine damage in the first letter of codons CGA and CGG, these codons are eliminated, leading to the overall dominance of codons CGU and CGC. This example can demonstrate the universality of the process of AT-enrichment of the genome, i.e. the universality of spontaneous mutations:

deamination of cytosine and oxidative guanine damage [1].

It would seem that the above is also valid for leucine codons: CUU, CUC, CUA, CUG, UUA, and UUG. However, this is not so for a number of reasons. Firstly, AT-enrichment of the genome can cause either emergence or elimination of codons CUA and CUG. Secondly, the mutation in codons CUU and CUC leads to the replacement of leucine by phenylalanine, and this is a much less significant event than the replacement of arginine by serine.

*Asymmetry in using adenine and uracil.*

Since the discovery of Chargaff's rules, it has been known that the amount of adenine in the DNA is equal to thymine. Within the mRNA structure, adenine and uracil are distributed asymmetrically, and sometimes this asymmetry is considerable.

In the database that we examined, just 23% of genomes contain larger amounts of uracil than adenine. These are mainly mitochondrial genomes as well as genomes of chloroplasts and some viruses. This asymmetry becomes much stronger if we consider only the first letter of the codon: having examined 4500 genomes, we can see that uracil prevails over adenine in the first letter in less than one hundred cases and these are only mitochondrial genomes. The most interesting result can be obtained by considering the third letter of the codon: uracil prevails in 80% cases, i.e. the result is opposite to the general trend. Moreover, adenine prevails over uracil in mitochondrial genomes! Thus, mitochondria become champions in nonrandom use of uracil in their codons. Both the revealed tendencies and the described "species specificity" can be accounted for by the chemical properties of uracil and the structure of the corresponding anticodons.

It is easier for uracil to be converted into its enol form than for other nucleotides, which makes its use somewhat ambiguous [2]. It can form pairs with both adenine and guanine, and this can lead to translation mistakes. Thus, it seems advantageous to limit, as much as possible, the use of uracil in the encoding part of the codon.

The third position of the codon is perfectly protected against such problems: the replacement of cytosine by uracil does not affect the interpretation. However, the replacement of guanine by adenine can cause significant changes. Therefore, in the third position of the codon, cytosine deamination is asymmetrically punished by selection. This mutation is harmless in the encoding sequence and significant in the complementary one.

Why are mitochondria a twofold exception? Both points of the above statement are wrong for mitochondrial codons. There is no difference between adenine and guanine in the third letter, and both uracil and cytosine are inconvenient, as during codon-anticodon interaction the uracil-uracil (or uracil-cytosine) pair has to be formed.

**Conclusions.**

All of the revealed tendencies ensue from the structure of the genetic code and the universality of spontaneous mutations. If the examination is restricted to the organisms that use the "universal genetic code", the revealed tendencies will grow into laws. Thus, the reading frame for the genomes using the universal genetic code can be found via a simple algorithm: "In the first position the amount of adenine is larger than the amount of uracil; in the third position the opposite is true".

It is worth noting that the basic reasons for the discussed tendencies are different and, clearly, they alone cannot account for the diversity of variants of codon usage.

**Acknowledgement**

The author would like to thank Krasova E. for her assistance in preparing this manuscript.

**References:**

1. Semenov D.A. Evolution of the genetic code. From the CG- to the CGUA-alphabet, from RNA double helix to DNA. arXiv:0805.0484

2. Semenov D.A. Wobbling of What? arXiv:0808.1780